\newcommand{\ifrevtex}[2]{#2}

\ifrevtex{
  \documentclass[10pt,
  tightenlines,
  amsmath,amssymb,
  nofootinbib,
  prd,
  twocolumn,
  showpacs,
  preprintnumbers]{revtex4}
  }{
  \documentclass[hyper]{JHEP3}
  }

\usepackage{graphicx}
\usepackage{psfrag}
\usepackage{bm}

\ifrevtex{}{\usepackage[numbers]{natbib}}
\ifrevtex{}{\usepackage{amsmath}}
\ifrevtex{}{\usepackage{amssymb}}

\ifrevtex{
\begin{document}}{}

\newcommand{\Lagrangian}{\mathcal{L}}
\newcommand{\exclude}[1]{}

\ifrevtex{}{\newcommand{\widetext}{}}
\ifrevtex{}{\newcommand{\affiliation}[1]{}}
\ifrevtex{}{\newcommand{\pacs}[1]{}}
%\ifrevtex{}{\newcommand{\acknowledgments}{\section{Acknowledgements}}}
\ifrevtex{\newcommand{\abs}[1]{\begin{abstract}#1\end{abstract}}
  }{
  \newcommand{\abs}[1]{\abstract{#1}}
  }
\ifrevtex{\newcommand{\fig}[1]{
    \begin{figure}[ht]
      \begin{center}
        #1
      \end{center}
    \end{figure}}
  }{
  \newcommand{\fig}[1]{\FIGURE[ht]{#1}}
  }

\title{Global Strings in High Density QCD}
\ifrevtex{
\author{Michael McNeil Forbes}
\affiliation{Center for Theoretical Physics, Massachusetts Institute
  of Technology, Cambridge, MA, 02139, USA}
\author{Ariel R. Zhitnitsky}
\affiliation{Department of Physics and Astronomy, University of British Columbia,
  Vancouver, BC V6T 1Z1, Canada}
}{
\author{Michael McNeil Forbes\\
  Center for Theoretical Physics, Massachusetts Institute
  of Technology, Cambridge, MA, 02139, USA}
\author{Ariel R. Zhitnitsky\\
  Department of Physics and Astronomy, University of British Columbia,
  Vancouver, BC V6T 1Z1, Canada}
}
\date{\today}

\preprint{hep-ph/0109173, MIT-CTP-3190}
\abs{
  We show that several types of global strings occur in
  colour superconducting quark matter due to the spontaneous violation
  of relevant $U(1)$ symmetries.  These include the baryon $U(1)_B$,
  and approximate axial $U(1)_A$ symmetries as well as an approximate
  $U(1)_S$ arising from kaon condensation.  We discuss some general
  properties of these strings and their interactions.  In particular,
  we demonstrate that the $U(1)_A$ strings behave as superconducting
  strings.  We draw some parallels between these strings and global
  cosmological strings and discuss some possible implications of these
  strings to the physics in neutron star cores.
} \pacs{Not Specified.}

\ifrevtex{}{\begin{document}}

\ifrevtex{\maketitle}{}

\section{Introduction}
\label{sec:introduction}
Domain walls and strings are common examples of topological defects
which are present in various field
theories~\cite{Rajaraman:1982,Vilenkin:1994}.  Domain walls are
configurations of fields related to the nontrivial mapping $\pi_0(M)$,
while topologically stable strings are due to the mapping
$\pi_{1}(M)$ where $M$ is manifold of degenerate vacuum states.  While
it is generally believed that there are no domain walls or strings in
the Standard Model due to the triviality of the corresponding
mappings, such objects may exist in extensions to the Standard Model,
or in phases where the symmetries of the Standard Model are broken.

Topological defects that occur in extensions to the Standard Model may
play an important role in cosmology as described in the
book~\cite{Vilenkin:1994}, however, the focus of this paper will be
strings and domain walls that exist in high density matter where the
symmetries of the Standard Model are broken.  As was recently
demonstrated~\cite{Son:2000fh}, in the regime of high baryon densities
when the chemical potential $\mu$ is much larger than the QCD scale
$\mu\gg \Lambda_{\textrm{QCD}}$, QCD supports domain walls.  The existence and
long life time of these domain walls is based on the following facts:
1) the instanton density is small at large chemical potential,
suppressing the effect of the chiral anomaly and giving rise to the
approximate $U(1)_A$ symmetry; 2) this $U(1)_A$ symmetry is
spontaneously broken, and 3) the decay constant of the pseudoscalar
singlet boson ($\eta'$) is large and its mass small at large $\mu$.

In this paper we show that, along with these domain walls, there exist
topological string configurations with interesting physical
properties.  Specifically, we discuss the three flavour $N_f=3$
colour-flavour locking phase (CFL), and the two flavour $N_f=2$
superconducting phase (2SC). In the $N_f=3$ CFL phase both the
$U(1)_B$ and the approximate $U(1)_A$ symmetries of QCD are
spontaneously broken.  In addition, a third approximate $U(1)_S$
symmetry, related to the formation of a kaon condensate, is broken
(see Section~\ref{sec:cfl-phase} below).
Thus, at least three types of 
strings are possible in the presence of a
kaon condensate and two types
of strings are possible if no kaon condensate forms.  In the $N_f=2$
2SC phase, the $U(1)_B$ symmetry is not broken and we only expect one
type of string from the breaking of the approximate $U(1)_A$ symmetry.
The important point is that the asymptotic freedom of QCD allows us to
assert the existence of these strings in the high baryon density
regime, and the properties of these strings can be determined by
controllable weak-coupling calculations.

Na\"\i{}vely, these strings are neutral objects with respect to
electromagnetic interactions, and therefore, one might think that they
play no role in electromagnetic dynamics.  However, for the $U(1)_A$
strings, this na\"\i{}ve expectation is incorrect due to the anomalous
coupling of the singlet Goldstone field with a massless
``photon''---actually a mixture of the bare electromagnetic field and
one component of the colour gluon fields.  We demonstrate that such
strings behave as superconducting strings, and we estimate the maximum
current that they can carry.  Thus, QCD strings at high density may
affect the electromagnetic properties of high density matter,
influencing, for example, the magnetic fields in the cores of neutron
stars.

Regarding the baryonic $U(1)_B$ strings present in the CFL phase: it
has been known for quite a while~\cite{Davis:1989gn} that, although
superfluid vortices (in liquid helium for example) and static global 
string solutions (related to the spontaneous breaking of a global
symmetry, $U(1)_B$ in this case) are closely related, there are some
differences.  In particular, fluid vortices carries angular momentum
while a static global strings have no angular momentum.
What was demonstrated in~\cite{Davis:1989gn} was that the two can be
identified if the static strings are immersed in a uniform background
with nonzero density.  In high density QCD, such a background is
naturally present, thus we demonstrate that the $U(1)_B$ global
strings can be identified with the superfluid vortices likely to form
in a rotating CFL phase with nonzero angular momentum as has been
conjecture to exist in the cores of neutron stars (see for
example~\cite{Alford:2000sx,Rajagopal:2000uu}).  A similar mechanism
leads to the formation of $U(1)_S$ strings.

The mechanism that forms the $U(1)_A$ strings in the $2SC$ phase,
which do not carry angular momentum, is less obvious to us and, in
this paper, we simply assume that some mechanism exists to form these
types of objects.  In general, there will be complicated dynamical
interactions between the strings, which may lead to the formation of
more complicated stable objects like rings, springs and vortons.  In
this paper we limit ourselves to describing the properties of
individual strings and their dominant  pairwise
interactions.

This paper is organized as follows: In Section~\ref{sec:vortices} we
describe the string solutions and calculate the relevant parameters of
the string in terms of the fundamental parameters of the theory.  In
Section~\ref{sec:spinning-strings} we discuss how the $U(1)_B$ strings
present in the CFL phase are not static, but rather spin with a time
dependent phase.  These spinning strings can then be identified with the
superfluid vortices.  In Section~\ref{sec:electr-prop} we discuss the
anomalous electromagnetic properties of the $U(1)_A$ strings.  We
solve the relevant Maxwell equations in the string background and
demonstrate that the $U(1)_A$ strings behave as if they are
superconducting.  Finally, in the Conclusion~\ref{sec:conclusion}, we
discuss the interactions between strings and speculate on their
effects on the physics of high density quark matter and neutron stars.

\section{Strings in High Density QCD.}
\label{sec:vortices}
When one considers QCD at high density, the relevant excitations are
due to quarks with momenta close to the Fermi surface.  At high
densities, these low-energy excitations have high momenta and, because
of asymptotic freedom, one might hope to regain theoretical control in
a weak coupling regime.  Although the situation is not quite as simple
as it appears, we do gain theoretical control.  In particular, the
colour anti-triplet $\mathbf{\bar{3}}$ single gluon interaction is
attractive, and, for low energies, this leads to a
preferred Bardeen-Cooper-Schrieffer (BCS) pairing of quarks with opposite
momenta that reduces the energy of the vacuum state.  This is the
phenomena of colour superconductivity.  (For a nice review, see~\cite{Rajagopal:2000wf}.)

To be specific, we consider the simplest model where high-density
$U(1)_A$ type strings appear: QCD with $N_f=2$ massless quark flavours
($u$ and $d$) and $N_c=3$ colours.  This model is a rather good
approximation to realistic quark matter at moderate densities.  At
higher densities, the approximation of $N_f=3$ massless quarks becomes
quit good.  This will be discussed later in
Section~\ref{sec:cfl-phase}.  The most important qualitative
difference between the $N_f=2$ and the $N_f=3$ phase is the emergence
of new spontaneously broken symmetries: the $U(1)_B$ in the three
flavour case and the $U(1)_S$ symmetry if a kaon condensate forms in
the CFL phase.  As a consequence, if a kaon condensate forms in the
CFL phase, then there emerge two new types of global strings related
to the $U(1)_B$ and $U(1)_S$ symmetries.  If a kaon condensate does
not form, then only one new type of string related to the $U(1)_B$
symmetry emerges.  In what follows we use the same normalization
factors as the paper~\cite{Son:2000fh} on domain walls in dense QCD.

\subsection{$N_f=2$ Superconducting Phase (2SC)}
\label{sec:2SC}
We recall that the ground state at high baryon densities is a
superconducting
state~\cite{Barrois:1977xd,Frautschi:1978,Barrois:1979,Bailin:1984bm,Alford:1998zt,Rapp:1998zu},
characterized by the condensation of diquark Cooper pairs.  The
superconducting ground state spontaneously breaks the symmetry of the
bare QCD Lagrangian through the non-zero diquark condensates $\langle
\Psi_{\alpha}^{ia}\Psi_{\beta}^{jb}\rangle$ which represent the Cooper
pairs.  Here we explicitly show the flavour ($\alpha$, $\beta$, etc.),
colour ($a$, $b$, etc.) and spinor ($i$, $j$, etc.)  indices.  In this
section we consider the $N_f=2$ case where the strange quark is
treated as heavy.  In this 2SC phase, the diquark condensates have the
form:
\begin{subequations}
  \label{eq:2SC}
  \begin{align}
    \langle \Psi_{La}^{i\alpha}\Psi_{Lb}^{j\beta}\rangle^*
    &= \epsilon^{ij}\epsilon^{\alpha\beta}\epsilon_{abc}X^c,\\
    \langle \Psi_{Ra}^{i\alpha}\Psi_{Rb}^{j\beta}\rangle^* 
    &= \epsilon^{ij}\epsilon^{\alpha\beta}\epsilon_{abc}Y^c.    
  \end{align}
\end{subequations}
The condensates $X^c$ and $Y^c$ are complex colour 3-vectors which are
aligned along the same direction in the ground state.  They
spontaneously break the colour $SU(3)_c$ group down to $SU(2)_c$.  The
lengths of these vectors are equal $|X|=|Y|$ and have been computed
perturbatively~\cite{Son:2000fh,Son:1998uk}:
\begin{equation}
  \label{eq:cond_mag}
  |X|=|Y|=\frac{3}{2\sqrt{2}\pi}\frac{\mu^2\Delta}{g}.
\end{equation}
In perturbation theory, there is an approximate degeneracy of the
ground state with respect to the relative $U(1)_A$ phase between $X^c$
and $Y^c$ which is a symmetry of the QCD Lagrangian at the classical
level.  A nonzero value~(\ref{eq:2SC}) for the vacuum condensate
implies that the $U(1)_A$ symmetry is spontaneously broken and, thus,
the corresponding pseudo-Goldstone boson---the $\eta'$---enters into
the theory.  The $U(1)_B$ symmetry of the QCD Lagrangian also appears
to be broken, but is in fact restored by a simultaneous $SU(3)_c$
rotation.  Thus, only the $U(1)_A$ symmetry is spontaneously broken in
the $N_f=2$ 2SC phase.

Following~\cite{Son:2000fh}, to describe the $\eta'$ physics in an
explicit way, we construct the gauge-invariant order parameter
\begin{equation}
  \Sigma=Y^\dagger X \equiv Y_c^*X^c,
\end{equation}
where the phase $\varphi_A$ is to be identified with dynamical $\eta'$
field,
\begin{equation}
  \Sigma = |\Sigma|e^{-i\varphi_A} = |\Sigma|e^{-i\eta'/f}  . 
\end{equation}
It is evident that $\Sigma$ carries a non-zero $U(1)_A$ charge:
\begin{subequations}
  \label{eq:Atransformation}
  \begin{align}
    \Psi \rightarrow e^{i\gamma_5\alpha/2}\Psi 
    &\Rightarrow 
    X \rightarrow e^{-i\alpha},\text{ and }
    Y \rightarrow e^{i\alpha}, \\
    &\Rightarrow
    \Sigma \rightarrow e^{-2i\alpha},\\
    &\Rightarrow
    \varphi_A \rightarrow \varphi_A+2\alpha.
  \end{align}
\end{subequations}
At low energies, the dynamics of the Goldstone mode $\varphi_A$ are
described by an effective Lagrangian, which, to leading order in
derivatives, must take the following form~\cite{Son:2000fh},
\begin{equation}
  \label{eq:Leff}
  \Lagrangian = f^2 [(\partial_0\varphi_A)^2 - u^2 (\partial_i\varphi_A)^2] -
  V_{\mathrm{inst}}(\varphi_A).
\end{equation}
For large chemical potentials $\mu\gg\Lambda_{\mathrm{QCD}}$, the
leading perturbative values for the decay constant $f$ and velocity
$u$~\cite{Beane:2000ms}, and the instanton contribution describing the
explicit anomalous breaking of the $U(1)_{\mathrm{A}}$
symmetry~\cite{Son:2000fh} have been calculated:
\begin{subequations}
  \label{eq:Vinst}
  \begin{gather}
    V_{\mathrm{inst}}(\varphi_A) = -a \mu^2\Delta^2\cos\varphi_A,\\
    \begin{aligned}
      f^2 &= \frac{\mu^2}{8\pi^2}, & u^2 &= \frac{1}{3}.
    \end{aligned}
  \end{gather}
\end{subequations}
In this formula $\Delta$ is the BCS gap, and
$a$ is a dimensionless parameter~\cite{Son:2000fh}
\begin{equation}
  \label{eq:a}
  a \sim   \left(\frac{\Lambda_{\mathrm{QCD}}}{\mu} \right)^{29/3}
\end{equation}
that goes to zero at large $\mu$.  In what follows it will be
important that the $\eta'$ mass is asymptotically small, as can be seen
from~(\ref{eq:Leff}):
\begin{equation}
  \label{eq:meta}
   m_{\eta'} = \frac{\mu}{f} \Delta \sqrt{\frac{a}{2}}
   = 2\pi \Delta  \sqrt a\, .
\end{equation}

The effective Lagrangian~(\ref{eq:Leff}) is justified for describing
the light $\eta'$ degree of freedom, but to describe global strings we
must formulate an effective theory for fluctuations in the magnitude
of the condensate $|\Sigma|$.  From~(\ref{eq:cond_mag}) we have that
$|\langle\Sigma\rangle| = 9\mu^4\Delta^2/(8\pi^2g^2)$.  Thus, we
introduce the dynamical field $\Phi(x)$ of dimension one and
expectation value $|\langle\Phi\rangle| = \langle\rho\rangle=\Delta$:
\begin{equation}
  \label{eq:phidef}
  \Sigma(x) = \Delta\left(\frac{3\mu^2}{2\sqrt{2}\pi
 g}\right)^2\Phi(x).
\end{equation}
With this definition, the $\eta'$ dynamical field is merely the phase
of a complex field
\begin{equation}
  \Phi = |\Phi|\exp(-i\varphi_A) = \rho \exp(-i\varphi_A)
\end{equation}
as in the Abelian Higgs model.

The effective potential for this condensate has been
calculated~\cite{Miransky:1999tr} for asymptotically large $\mu$.  In
terms of the field $\rho(x) = |\Phi(x)|$, an approximate
expression\footnote{Note, the normalization for the condensate
  in~\cite{Miransky:1999tr} differs from our normalization for $|X|$
  by a factor of $16$.} for the potential~\cite{Miransky:1999tr} can
be used which is a good description for $\rho$ close to its vacuum
expectation value $\rho\approx \langle\rho\rangle = \Delta$
\begin{equation}
  \label{eq:Veff}
  V_{\Phi}(\rho)=-\frac{\mu^2\Delta}{\pi^2}
   \rho\left[1-\ln \left(\frac{\rho}{\Delta}\right)\right].
\end{equation}
Finally, in terms of a single complex field $\Phi=\rho \exp
(-i\varphi_A)$ the effective Lagrangian describing the $\eta'$ phase
$\varphi_A$ and the absolute value for the condensate $\rho$ can be
represented in the following simple way:
\ifrevtex{
  \begin{widetext}
    \begin{subequations}
      \label{eq:LeffString}
      \begin{align}    
        \Lagrangian_{\mathrm{eff}} &=
        \frac{f^2}{\Delta^2}\left(|\partial_0\Phi|^2
          -u^2|\partial_i\Phi|^2\right) 
        +\frac{\mu^2\Delta}{\pi^2}
        \rho \left[1-\ln\left(\frac{\rho}{\Delta}\right)\right]
        +a\mu^2\Delta\rho\cos\varphi_A\\
        &=\frac{f^2}{\Delta^2}\left((\partial_0\rho)^2
          -u^2(\partial_i\rho)^2 \right)
        +\frac{f^2\rho^2}{\Delta^2}\left((\partial_0\varphi_A)^2
          -u^2(\partial_i\varphi_A)^2 \right)
        +\frac{\mu^2\Delta}{\pi^2}
        \rho\left[1-\ln \left(\frac{\rho}{\Delta}\right)\right]
        +a\mu^2\Delta\rho\cos\varphi_A
      \end{align}
    \end{subequations}
  \end{widetext}
  }{
  \label{eq:LeffString}
  \begin{align}    
    \Lagrangian_{\mathrm{eff}} &=
    \frac{f^2}{\Delta^2}\left(|\partial_0\Phi|^2
      -u^2|\partial_i\Phi|^2\right) 
    +\frac{\mu^2\Delta}{\pi^2}
    \rho \left[1-\ln\left(\frac{\rho}{\Delta}\right)\right]
    +a\mu^2\Delta\rho\cos\varphi_A\nonumber\\
    \begin{split}
      =\frac{f^2}{\Delta^2}\left((\partial_0\rho)^2
        -u^2(\partial_i\rho)^2 \right)
      +\frac{f^2\rho^2}{\Delta^2}\left((\partial_0\varphi_A)^2
        -u^2(\partial_i\varphi_A)^2 \right)+\\
      +\frac{\mu^2\Delta}{\pi^2}
      \rho\left[1-\ln \left(\frac{\rho}{\Delta}\right)\right]
      +a\mu^2\Delta\rho\cos\varphi_A
    \end{split}
  \end{align}
  }
where the normalization factor $f^2/\Delta^2$ for the kinetic term has
been chosen to reproduce~(\ref{eq:Leff}) where
$\rho\sim\langle\rho\rangle$ which correctly describes dynamics of the
light pseudo-Goldstone $\eta'$ meson.

We should comment here that the potential presented in~(\ref{eq:Veff})
was derived in~\cite{Miransky:1999tr} for very large $\mu$ and when
the field $\Phi$ is close to its vacuum expectation value
$\langle\rho\rangle\simeq\Delta$.  It deviates considerably from this
form when $\rho$ is far from its vacuum expectation value
$\rho\neq\Delta$.  Besides that, there is no justification to keep
only the lowest derivative term for the massive mode $\rho$ in the
expression ~(\ref{eq:LeffString}).  Finally, there is an ambiguity in
the definition of the dynamical field $\rho$ describing heavy $\sim
\Delta$ degrees of freedom: any smooth function of $F(\rho)$ (for
example $\exp(\rho)$) is appropriate for the description of the
dynamics of the heavy degrees of freedom. This is a marked contrast
with description of the light Goldstone fields where the physics does
not depend on specific parameterization of the light fields.  There
are many other deficiencies of the Lagrangian~(\ref{eq:LeffString})
describing massive field $\rho$ which we shall not comment about.

We are not pretending to have derived a Lagrangian describing a heavy
(order $\Delta$) degree of freedom. Rather, we want to demonstrate the
qualitative features of the effective potential: that it has a Mexican
hat shape (as it should) and that the internal relevant scales are of
order $\Delta$ and not $\mu$. Indeed, the general scale $\sim\mu^2$
factors from the expression~(\ref{eq:LeffString}).  Therefore, the
description of the strings which follows serves mainly to illustrate
the qualitative features of the strings.  In particular, the details
of the strings core are not well founded, however, we shall see that
the behaviour of the strings far from the core is governed by the form
of the potential where $\rho\sim\langle\rho\rangle=\Delta$ where the
effective potential is valid.  This region gives a logarithmic
contribution to the string tension which dominates the energy of the
strings.  Thus, despite the fact that the effective
theory~(\ref{eq:LeffString}) breaks down for small $\rho$, the
description of the strings far from the core is well justified.

\subsubsection{Global Strings}
\label{sec:global-strings}
First we consider the properties of an isolated global string which is
symmetric about the $z$ axis.  The term
$V_{\mathrm{inst}}(\varphi_A)$, which explicitly breaks the $U(1)_A$
symmetry, is small, and vanishes in the high density limit.  Thus, our
approximation of the global string is a good description out to
lengths scales of order $m_{\eta'}^{-1}$.

With this simplification, we are looking for a static, classical field
configuration $\Phi(r,\phi) =
\rho(r,\phi)\exp(-i\varphi_A(r,\phi))$ which minimizes the energy
density or string tension $\alpha$:
\begin{equation}
  \alpha = \iint\Bigl(\mathcal{H}(\rho,\varphi_A)
    -\mathcal{H}_\mathrm{vac}\Bigr)\;\mathrm{d}{x}\mathrm{d}{y}
\end{equation}
where $\mathcal{H}$ is the energy density of the field configuration:
\begin{equation*}
  \mathcal{H} = \frac{f^2u^2}{\Delta^2}\left(
    (\partial_i\rho)^2 
    +\rho^2(\partial_i\varphi_A)^2
    -\frac{\mu^2\Delta^3}{\pi^2f^2u^2}
    \rho\left[1-\ln \left(\frac{\rho}{\Delta}\right)\right]
  \right),
\end{equation*}
and $\mathcal{H}_\mathrm{vac} =
\mathcal{H}(\langle\rho\rangle,\langle\varphi_A\rangle) =
-\mu^2\Delta^2/\pi^2$ is a trivial shift of the background vacuum
energy introduced so that $\mathcal{H}\to 0$ far from the core of the
string.

To simplify the equations, we introduce the dimensionless field
$\tilde{\rho}$ and the dimensionless coordinates $\tilde{x}_i$:
\begin{align}
  \tilde{\rho} &= \frac{\rho}{\Delta}, &
  \langle\tilde{\rho}\rangle &= 1, &
  \tilde{x}_i &= mx_i = 2\sqrt{3}\Delta x_i.
\end{align}
The natural length scale is thus set by the parameter
$m=2\sqrt{3}\Delta$: the mass of the excitations about the condensate
$\rho=\langle\rho\rangle+\delta\rho$ in our model~(\ref{eq:Veff}).  In
terms of these dimensionless parameters, the energy density becomes
\begin{align*}
  \mathcal{H} &= 24f^2u^2\Delta^2 \Bigl(
    \tfrac{1}{2}(\partial_i\tilde{\rho})^2 
    +\tfrac{1}{2}\tilde{\rho}^2(\partial_i\varphi_A)^2 
    -\tilde{\rho}\left[1-\ln \tilde{\rho}\right]
  \Bigr)
\end{align*}
where all the derivatives are with respect to the dimensionless
coordinates $\tilde{x}_i$.  To minimize the string tension, we can
drop the overall factor and then determine the equations of motion.
To achieve the appropriate boundary conditions, $\varphi_A$ will wind
uniformly $n$ times as a function of $\phi$
\begin{equation}
  \label{eq:n}
  \varphi_A(r,\phi) = \varphi_A(\phi) = n\phi 
  =n\tan^{-1}\left(\frac{y}{x}\right).
\end{equation}
Converting to polar coordinates we have
\begin{equation}
  \label{eq:tension}
  \alpha =4\pi f^2 u^2
  \int_0^{\tilde{R}} \left(\frac{\dot{\tilde{\rho}}^2}{2} 
  + \frac{\tilde{\rho}^2n^2}{2\tilde{r}^2}
  + V(\tilde{\rho}) \right)\tilde{r}\mathrm{d}\tilde{r}
\end{equation}
where the dot signifies differentiation with respect to $\tilde{r}$,
\begin{equation}
  V(\tilde{\rho}) =
  1-\tilde{\rho}\left[1-\ln\tilde{\rho}\right],
\end{equation}
and we have introduced an outer limit $\tilde{R}$ to the string's size
to make the tension finite.  The equations of motion follow from an
application of a variational principle:
\begin{equation}
  \ddot{\tilde{\rho}}+\frac{\dot{\tilde{\rho}}}{\tilde{r}}
  =\frac{\tilde{\rho} n^2}{\tilde{r}^2}
  +\frac{\mathrm{d}V(\tilde{\rho})}{\mathrm{d}\tilde{\rho}}
\end{equation}
with boundary conditions $\tilde{\rho}(0)=0$ and
$\tilde{\rho}(\infty)=1$.  The solution for the lowest energy string
with winding $n=1$ is presented in Figure~\ref{fig:string}.
\fig{
    \psfrag{r (in units of XXXX)}{$\tilde{r}$ (in units of $2\sqrt{3}\Delta$)}
    \psfrag{dimensionless amplitude}{Dimensionless amplitude.}
    \psfrag{p(r)}{$\tilde{\rho}(\tilde{r})$}
    \psfrag{dp(r)}{$\dot{\tilde{\rho}}(\tilde{r})$}
    \psfrag{ddp(r)}{$\ddot{\tilde{\rho}}(\tilde{r})$}
    \psfrag{String solution}{}
    
    \ifrevtex{\includegraphics[width=0.49\textwidth]{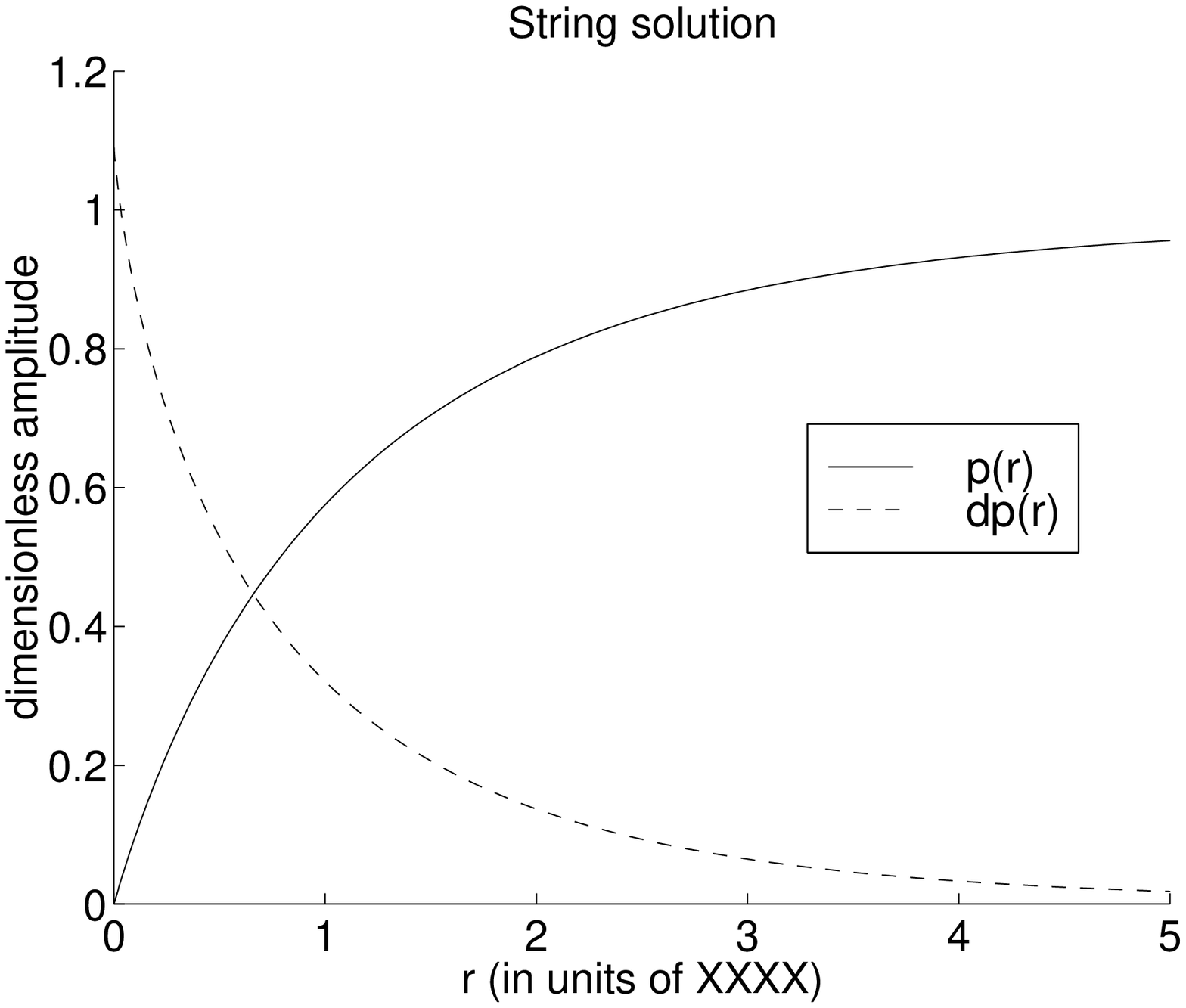}}
    {\includegraphics[width=0.8\textwidth]{string.eps}}
    
    \caption{
      \label{fig:string}
      Radial $\tilde{\rho}(\tilde{r})$ dependence of the $n=1$ string in
      dimensionless units: the vertical scale is set by $\Delta$ and the
      horizontal scale is set by $(2\sqrt{3}\Delta)^{-1}$.
      }
}

The string is governed by two relevant parameters: its core size and
its tension.  As can be seen in Figure~\ref{fig:string}, the core size
is of order $\tilde{r}_c\sim 1$ while the tension of a global string
diverges as $\log(\tilde{R})$ where $\tilde{R}$ is an upper cutoff
determined by the environment of the string.  For large distances the
second term in~(\ref{eq:tension}) dominates and we have a logarithmic
divergence.  We plot the cumulative energy as a function of the upper
cutoff $\tilde{R}$ in Figure~\ref{fig:energy}.

Physically, the effective potential~(\ref{eq:Veff}) is only valid for
$\rho\sim\langle\rho\rangle$.  Thus, the details of the string core
must be interpreted with caution.  The large distance behaviour and
the logarithmic divergence in the string tension, however, are well
justified.
\fig{
  \psfrag{Cumulative Energies}{Cumulative Energies}
  \psfrag{Outer string radius (in units of XXXXX)}
  {Outer string radius $\tilde{R}$ (in units of $(2\sqrt{3}\Delta)^{-1}$)}
  \psfrag{Energy (In units of XXXXX)}
  {Energy (in units of $4\pi f^2 u^2$)}
  \psfrag{dp*dp/2*r}
  {$\int_0^{\tilde{R}}\tilde{r}\dot{\tilde{\rho}}^2/2\mathrm{d}\tilde{r}$}
  \psfrag{p*p*n*n/r/2}
  {$\int_0^{\tilde{R}}\tilde{\rho}^2/2\tilde{r}\mathrm{d}\tilde{r}$}
  \psfrag{rV}
  {$\int_0^{\tilde{R}}\tilde{r}V\mathrm{d}\tilde{r}$}
  \ifrevtex{\includegraphics[width=0.49\textwidth]{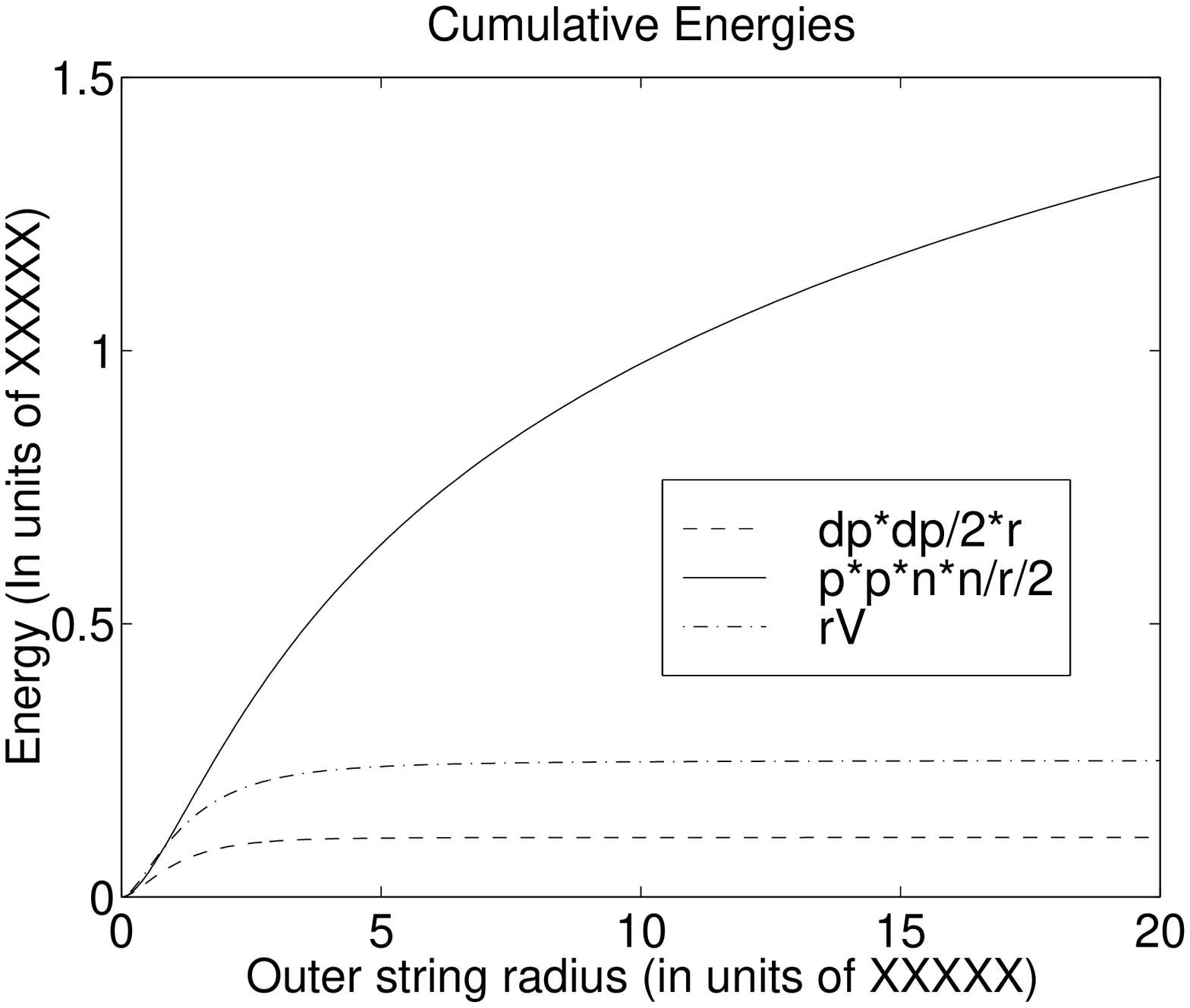}}
  {\includegraphics[width=0.8\textwidth]{energy.eps}}
  \caption{
    \label{fig:energy}
    Cumulative contributions to the string tension of an $n=1$ string
    from the core out to the cutoff radius $\tilde{R}$ in
    dimensionless units: the vertical scale is set by $4\pi f^2 u^2$
    and the horizontal scale is set by $(2\sqrt{3}\Delta)^{-1}$.  The
    three curves correspond to the three terms in~(\ref{eq:tension}).
    The first kinetic term approaches $0.11$ while the potential term
    approaches $0.25$.  Notice that the second term gives a logarithmic
    divergence in $\tilde{R}$.}
}

\subsubsection{Domain Walls}
\label{sec:domain-walls}
The formation of global strings discussed in the previous section
neglected the instanton contribution $V_{\text{inst}}(\varphi_A)$
which explicitly breaks the $U(1)_A$ symmetry responsible for the
global strings.  Thus, the preceding analysis and picture is really
only justified on distance scales small compared to those set by the
anisotropy $V_{\text{inst}}(\varphi_A)$, i.e. for $r \ll R\sim
m_{\eta'}^{-1}$, which, in the high density limit, is much larger than
the core size $\Delta$ as can be seen from~(\ref{eq:a}) and
(\ref{eq:meta}).  For distances larger than $r \gg R\sim
m_{\eta'}^{-1}$, the appropriate description is no longer one of
global strings, but one of QCD domain walls bounded by strings: the
situation is similar to the so-called $N=1$ axion
model~\cite{Vilenkin:1982ks}.

Here we summarize a few results regarding QCD domain walls that will
be relevant for our discussions later on.  As classically stable
objects, QCD domain walls were first analyzed in~\cite{Forbes:2000et}
for $\mu=0$ in the large $N_c$ limit.  Similar objects were later
shown to be stable for $N_c=3$ at high densities~\cite{Son:2000fh}.
We shall not repeat the analysis presented in~\cite{Son:2000fh} but
quote some important results that we will make use of later.

The thickness of the domain walls is set by the mass $m_{\eta'}$
(\ref{eq:meta}) of the excitations in the $\varphi_A$ field about the
true vacuum $\varphi_A=0$.  In addition, the energy density per unit
area of the domain walls (the wall tension) is
\begin{equation}
  \label{eq:DWtension}
  \sigma = 8\sqrt{2a} uf\mu\Delta \sim
  \sqrt{a} \mu^2 \Delta.
\end{equation}

Thus, on scales much smaller than $R\sim m_{\eta'}^{-1}$, the
description of the global strings is valid, however, when one looks at
scales larger than this, then one sees that these strings are really
attached to domain walls.

\subsection{$N_f=3$ CFL Phase}
\label{sec:cfl-phase}
For very high densities, it can be a good approximation to neglect the
strange quark mass.  In this case there will be three flavours $N_f=3$
in our model. Again, the ground state is a superconducting state,
characterized by the condensation of diquark Cooper
pairs~\cite{Alford:1998mk,Rapp:1999qa}.  The superconducting ground
state spontaneously breaks the symmetry of the bare QCD Lagrangian
through the non-zero diquark condensates $\langle
\Psi_{\alpha}^{ia}\Psi_{\beta}^{jb}\rangle$ which represent the Cooper
pairs.  (Again, we explicitly show the flavour ($\alpha$, $\beta$,
etc.), colour ($a$, $b$, etc.) and spinor ($i$, $j$, etc.) indices).
This condensate has the form
\begin{subequations}
  \label{eq:CFL}
  \begin{align}
    \langle \Psi_{La}^{i\alpha}\Psi_{Lb}^{j\beta}\rangle^* 
    &\sim \epsilon^{ij}\epsilon^{\alpha\beta \gamma}\epsilon_{abc}X^c_\gamma,\\
    \langle \Psi_{Ra}^{i\alpha}\Psi_{Rb}^{j\beta}\rangle^* 
    &\sim \epsilon^{ij}\epsilon^{\alpha\beta \gamma}\epsilon_{abc}Y^c_\gamma.    
  \end{align}
\end{subequations}
Now the condensates $X^c_\gamma$ and $Y^c_\gamma$ are complex
colour-flavour matrices.  Following~\cite{Son:1999cm,Son:2000tu}, we introduce
\begin{equation}
  \Sigma^\alpha_\beta = Y^\dagger X = (Y^*)^\alpha_c X^c_\beta.
\end{equation}
The matrix $\Sigma$ is a colour singlet and describes the meson octet
as well as the $\varphi_A$ axial singlet.  Unlike the $2SC$ phase, the
$U(1)_B$ symmetry is spontaneously broken giving rise to both axial
and baryonic strings.

To be precise, we consider the CFL phase and define the singlet phases
$\varphi_A$ and $\varphi_B$ describing the Goldstone bosons related to
the spontaneously broken symmetries $U(1)_A$ and $U(1)_B$ through the
following structure
\begin{subequations}
  \label{eq:CFL2}
  \begin{align}
    \langle \Psi_{La}^{i\alpha}\Psi_{Lb}^{j\beta}\rangle^* 
    &\sim \epsilon^{ij}\epsilon^{\alpha\beta c}\epsilon_{abc}
    e^{-i\varphi_A- i\varphi_B},\\
    \langle \Psi_{Ra}^{i\alpha}\Psi_{Rb}^{j\beta}\rangle^* 
    &\sim \epsilon^{ij}\epsilon^{\alpha\beta c}\epsilon_{abc}
    e^{i\varphi_A- i\varphi_B}.    
  \end{align}
\end{subequations}
In terms of the $\varphi_A$ and $\varphi_B$ fields, the corresponding
strings can be described as was done above for 2SC phase: the only
difference is that the $U(1)_B$ strings are not attached to domain
walls.  This is because the $U(1)_B$ symmetry is not explicitly broken:
i.e. the equations describing the $\varphi_B$ string contain no terms
analogous to those proportional to $a$ in (\ref{eq:LeffString}).

It was recently
argued~\cite{Schafer:2000ew,Bedaque:2001je,Kaplan:2001qk} that the
so-called ``symmetric CFL phase''---a pure form of the CFL phase where
no other meson condensates are formed---is unlikely to occur in
nature. Instead, it is likely that, in the CFL phase at high baryon
density, a kaon condensate forms~\cite{Schafer:2000ew,Bedaque:2001je,Kaplan:2001qk}. 
If this is the case,   the kaon
condensation breaks a new global $U(1)_S$ symmetry along with the
$U(1)_A$ and $U(1)_B$ symmetries discussed above. From Goldstone's
theorem one expects a new Goldstone boson $\varphi_S$, which is the
phase of the condensate, to appear in the spectrum.  One can present
the same arguments as before to deduce that a new type of the global
string related to $\varphi_S$ phase exists.  We shall call this
string the $U(1)_S$ string.  The equation describing this new $U(1)_S$
string is the same as (\ref{eq:LeffString}) with different
values $f_S$ and $a_S$ for the parameter $f$ and coefficient $a$ in
the Lagrangian (\ref{eq:LeffString}).  What is important is that, if
the decay constant $f_S \sim \mu$ is roughly the same as before and
proportional to $\mu$, then the coefficient $a_S$ is extremely small
(but not zero) as was argued recently~\cite{Son:2001xd},
\begin{align}
  \label{eq:K_0}
  a_S &\approx \frac{18\sqrt{2}}{g^2}G_F m_u m_s\cos\theta_C\sin\theta_C, &
  m_{\varphi_S}^2 &\sim a_S\Delta^2.
\end{align}
In this formula $\theta_C$ is Cabibbo angle and $G_F$ is the Fermi
constant.  Numerically, the scale corresponding to the
pseudo-Goldstone boson mass $m_{\varphi_S}$ is of order $m_{\varphi_S}\sim 50$
KeV.

\section{Spinning Strings}
\label{sec:spinning-strings}
The main goal of this section is to demonstrate the connection between
the idea of a global $U(1)_B$ string and the superfluid $U(1)_B$
rotational vortices in the CFL phase that carry angular momentum.
Since they carry angular momentum, they may be copiously formed in a
rotating neutron stars provided that the core of the star is in a
colour superconducting phase.

It is generally accepted that the global strings and non-relativistic
superfluid vortices are closely related; however, the behaviour of the
two are quite different. For example, near a straight superfluid
vortex at rest there is a velocity field moving circularly around it,
carrying momentum and kinetic energy.  But a simple global string (for
example defined by Lagrangian~(\ref{eq:Leff})) is a time independent
solution of the equations of motion. The momentum density away from
the string core is $\sim f^2\partial_0\varphi_A\partial_i\varphi_A$
which is zero in the rest frame of a static global string. The precise
relation between global strings which we discuss in this paper and
superfluid vortices was given in~\cite{Davis:1989gn}. There it was
demonstrated that, if a global string is immersed in a uniform,
Lorentz--non-invariant background, then the static solution (with
$\varphi_B$ to be identified with the azimuthal angle) is replaced by
a time-dependent solution $\varphi_B \rightarrow\varphi_B+ \omega t$
where the coefficient $\omega$ is determined by the density of the
background and the magnitude of the condensate.

In the quark superconducting phase, such a background is naturally
present and the constant $\omega$ is uniquely determined by the
chemical potential $\mu$.  The relevant Goldstone phases $\varphi_A$
and $\varphi_B$ are as defined in~(\ref{eq:CFL2}).  The effective
Lagrangian in the CFL phase for the axial Goldstone field $\varphi_A$
can be derived in the same manner as before~(\ref{eq:Leff})
\begin{equation}
  \label{eq:LeffA}
  \Lagrangian_A = f^2_A [(\partial_0\varphi_A)^2
  - u^2_A (\partial_i\varphi_A)^2]
  - V_{\mathrm{inst}}(\varphi_A) \,  .
\end{equation}
For large chemical potentials, $\mu\gg\Lambda_{\mathrm{QCD}}$, the
leading perturbative values for $f_A$ and $u_A$ have been
calculated~\cite{Son:1999cm,Son:2000tu,Zarembo:2000pj,Beane:2000ms}:
\begin{subequations}
  \begin{gather}  
    V_{\mathrm{inst}}(\phi_A) = -a'\mu^2\Delta^2\cos(\phi_A)\\
    \begin{aligned}
      f^2_A& = \frac{9\mu^2}{\pi^2},
      & u^2_A &= \frac{1}{3}.
      \label{eq:3}
    \end{aligned}
  \end{gather}
\end{subequations}
The instanton contribution that explicitly violates the $U(1)_A$
symmetry has also been calculated~\cite{Son:2001jm}
\begin{equation}
  \label{eq:a'}
  a' \sim
  \left(\frac{m_s}{\mu}\right)\left(\frac{\Lambda_{\mathrm{QCD}}}{\mu}
  \right)^{9}
\end{equation}
which again vanishes for large $\mu$.

To leading order in perturbation theory, the parameters $f_B$ and
$u_B$ for the Goldstone boson associated with the spontaneous breaking
of ${U(1)_B}$ symmetry are identical to the ones presented
in~(\ref{eq:3})~\cite{Son:1999cm,Son:2000tu,Beane:2000ms}.  However,
the structure of the effective Lagrangian for the baryon Goldstone
field $\varphi_B$ is a little bit different due to the explicit
presence of the chemical potential $\mu$ in the original QCD
Lagrangian.

In order to restore the dependence on $\mu$, one can use the following
trick which, in the present context, was originally suggested
in~\cite{Kogut:1999iv,Kogut:2000ek} and consequently has been used in
a number of
papers~\cite{Son:1999cm,Son:2000tu,Beane:2000ms,Son:2000xc,Son:2000by,Son:2000by2}.
The idea is to make use of the fact that $\mu$ enters the QCD
Lagrangian in the same way as the zeroth component of a gauge
potential:
\begin{subequations}
  \begin{align}
    \Lagrangian_{\Psi} &= 
    \bar{\Psi}\left(i\gamma^\nu\partial_\nu - m\right)\Psi
    + \mu\Psi^\dagger\Psi,\\
    &= \bar{\Psi}\left(
      i\gamma^\nu\partial_\nu +\mu\gamma^0 - m
    \right)\Psi.
  \end{align}
\end{subequations}
Therefore, one can formally promote the global $U(1)_B$ baryon
symmetry to a local one by introducing a gauge field $B_{\nu}=(B_0,
\vec{0})$ coupled to the baryon current with coupling constant $\mu$:
\begin{equation}
  \Lagrangian_{\mathrm{QCD}} = 
  \bar{\Psi}\left(i\gamma^\nu \nabla_\nu - m\right)\Psi
\end{equation}
where $\nabla_\nu = \partial_\nu+i\mu B_\nu$.
 
Indeed, under the ${U(1)_B}$ rotations we have:
\begin{subequations}
  \begin{align}
    \Psi &\rightarrow e^{i\alpha}\Psi,\\
    \varphi_B &\rightarrow \varphi_B + 2\alpha,\\
    B_{\nu} &\rightarrow B_{\nu}-\mu^{-1}\partial_{\nu}\alpha,
  \end{align}
\end{subequations}
which leave the microscopical QCD Lagrangian unchanged.  An effective
description must respect this symmetry, and therefore, in the
effective Lagrangian description of the baryon Goldstone mode
$\varphi_B$ (\ref{eq:CFL2}), one must replace the derivative
$\partial_\nu\varphi_a$ in (\ref{eq:LeffA}) by the covariant derivative $D_\nu
\varphi_B = \partial_\nu\varphi_B+2\mu B_\nu$.  In matter with uniform
density, we fix $B_0=1$ and $B_i=0$ so that
\begin{align}
  D_0\;\varphi_B &\equiv (\partial_0 \varphi_B +2\mu) &
  D_i\;\varphi_B&\equiv \partial_i\varphi_B
\end{align}
and thus
\begin{equation}
  \label{eq:LeffB}
  \Lagrangian_B = f^2_B \bigl((D_0\varphi_B)^2 - u^2_B
  (D_i\varphi_B)^2\bigr).
\end{equation}
From this equation we see that we can restore the original form
(without covariant derivatives) if the baryonic Goldstone mode
$\varphi_B$---the phase of the condensate (\ref{eq:CFL2})---receives a
time dependence in the ground state
\begin{equation}
  \label{eq:Bw}
  \varphi_B \sim e^{i 2\mu t}.
\end{equation}
Note, in the previous literature concerning the CFL
phase~\cite{Beane:2000ms,Son:1999cm,Son:2000tu}, this time dependence
was irrelevant.  However, for the discussions of strings as
emphasized in \cite{Davis:1989gn} this dependence is essential.  In
particular, following~\cite{Davis:1989gn}, to analyze QCD strings and
their interactions in a nontrivial environment, one can introduce, an
equivalent description in terms of a two-index antisymmetric tensor
field $B_{\mu\nu}$ where
\begin{equation}
  f\partial_{\mu}\varphi_B 
  \sim\epsilon_{\mu\nu\lambda\sigma}\partial^{\nu}B^{\lambda\sigma}.
\end{equation}
Using this formalism, one can calculate the Lorentz force between
strings, similar to the Magnus force in the non-relativistic limit.
One can also introduce a quantity similar to the non-relativistic
vorticity and demonstrate that it is quantized (in our notations) in
units of $\mu^{-1}$.  We shall not discuss these interesting topics in
this paper: once the appropriate correspondence is made, the
techniques of~\cite{Davis:1989gn} can be applied.  The important
remark that we make is that, if the CFL phase with the global $U(1)_B$
symmetry is realized in the interior of a rotating neutron star where
there is a non-zero chemical potential $\mu$, then the global strings
that form will be spinning and will carry angular momentum.  Thus,
drawing upon the analogy with liquid helium, we expect that, if the
CFL phase rotates, then spinning global $U(1)_B$ strings will form.

One consequence of this connection between angular momentum and the
global string is that, if global strings are formed through the
transfer of angular momentum, then there will be a correlation between
the direction of the angular momentum and the sign of the winding of
the string.  To see this, note that the angular momentum of the field
$\varphi_B$ is proportional to:
\begin{equation}
  \label{eq:AngularM}
  M_{ij} \sim \int x_j\partial_0\varphi_B \partial_i\varphi_B
  - x_i\partial_0\varphi_B \partial_j\varphi_B.
\end{equation}
The terms $\partial_0\varphi_B \sim \mu$ pickup the sign of the
chemical potential and the terms $\partial_i\varphi_B \sim n$ pickup
the sign of the winding or topological ``charge'' $n$ (\ref{eq:n}) of
the string.  In the core of a neutron star, for example, the angular
momentum has a definite sign (as set by the rotation of the star) and
the chemical potential has a definite sign (the core is composed of
baryons, not antibaryons).  Thus, the sign of the topological charge
is correlated with the sign of the angular momentum.  If the formation
of these strings is related to the rotation of the bulk phase, then there
will be an excess of one type of string (either positive or negative
winding).

Exactly the same arguments can be presented for the $U(1)_S$ strings
mentioned at the end of the Section \ref{sec:cfl-phase} which
originate from kaon condensation.  The only modification that must be
made to Equation (\ref{eq:LeffB}) is to replace the chemical potential
$\mu$ by an induced chemical potential $\mu_{\text{eff}}\simeq
m_s^2/2p_F$, see~\cite{Schafer:2000ew,Bedaque:2001je,Kaplan:2001qk}.
The argument based on (\ref{eq:AngularM}) about the correlation
between the sign of the topological charge of the string with the sign
of the angular momentum, also holds.

The situation with $U(1)_A$ strings is expected to be quite
different: conservation of $P$ parity implies that the number of
strings and anti-strings must be the same.  This is due to the fact
that $U(1)_B$ strings do not transform into anti-strings under the
exchange $R\leftrightarrow L$ while $U(1)_A$ strings do, as can be
seen\footnote{We thank Misha Stephanov
for presenting this argument to us.} from the definition (\ref{eq:CFL2}).

\section{Electromagnetic Properties}
\label{sec:electr-prop}
Up until this point, we have discussed the existence of global $U(1)$
strings and their correspondence with superfluid vortices.  However,
since they only involve excitations close to the Fermi surface, there
will not be enough to affect the thermodynamics of the superconducting
phases.  In addition, the fields $\varphi_A$ and $\varphi_B$ are
neutral, so one might na\"\i{}vely expect that they have little effect
on electromagnetic physics either. It turns out, however, that the
axial strings $\varphi_A$ have non-trivial electromagnetic properties.

\subsection{Anomalous Electromagnetism of the $\eta'$}
\label{sec:anom-inter-eta}
In this section we are mainly concern with electromagnetic interaction
of the neutral $\eta'$ meson. The simplest way to derive the
corresponding low-energy effective Lagrangian, which includes the
massless electromagnetic $F_{\mu\nu}$ field and the light $\eta'$
field~(\ref{eq:Leff}), is to follow the standard procedure and consider
the transformation properties of the path integral under the
${U(1)_A}$ chiral transformation~(\ref{eq:Atransformation}).  As is known,
the measure is not invariant under these transformations due to the
chiral anomaly: it receives an additional contribution $\delta
\Lagrangian=(\alpha/2) \partial^{\nu}J_{\nu}^{A}$.  The expression for the
anomaly $\partial^{\nu}J_{\nu}^{A}$ is well known and takes the form:
\begin{equation}
  \label{eq:anomaly}
  \partial^{\nu}J_{\nu}^{A}
  =\frac{g^2}{16\pi^2} N_f G^a\tilde{G}^a
  +\frac{e^2}{8\pi^2} N_c F\tilde{F}\sum_{f=1}^{N_f}Q^2_{f} ,
\end{equation}
where we have included the electromagnetic fields $F_{\mu\nu}$ along
with the gluon fields $G_{\mu\nu}^a$ and their duals:
\begin{align}
  \tilde{F}_{\mu\nu} &= \frac{1}{2}
  \epsilon_{\mu\nu\lambda\sigma}F^{\lambda\sigma}, &
  \tilde{G}^a_{\mu\nu} &= \frac{1}{2}
  \epsilon_{\mu\nu\lambda\sigma}G^{a\lambda\sigma}.
\end{align}
One should note that the expression for the anomaly is an operator
relation which is valid for any finite chemical potential $\mu$.
Indeed, the anomaly arises from the ultraviolet properties of the
theory and is not sensitive to the finite chemical potential as long
as the regulator fields are heavier than $\mu$.

The second step is an identification of the parameters from the
Lagrangian $\delta L$ with the physical fields of the theory.  For
example, in QCD with $\mu=0$ the electromagnetic field $F_{\mu\nu}$
(\ref{eq:anomaly}) from the original theory is the observable physical
field (in contrast with superconducting phases when $\mu\neq 0$, see
below).  The parameter $\alpha$ of $U(1)_{\mathrm{A}}$ chiral
transformation is identified with the physical $\eta'$ field which is
nothing but the singlet phase of the chiral condensate,
\begin{equation}
  \langle\bar{\Psi}_R\Psi_L\rangle 
  \sim e^{-i\alpha} 
  \sim \exp\left(-i\frac{2\eta'}{f_{\pi}\sqrt{N_f}}\right).
\end{equation}
Therefore, the anomalous effective Lagrangian of $\eta'$ coupled to
photons takes a familiar form: 
\begin{equation}
  \delta \Lagrangian_{\eta'\gamma\gamma} = \frac{e^2}{8\pi^2}N_c
  F\tilde{F}\sum_{f=1}^{N_f}Q^2_{f}
  \frac{\eta'}{f_{\pi}\sqrt{N_f}}.
\end{equation}
In a similar manner, one can derive the well-known effective
Lagrangian describing the famous $\pi^0 \to \gamma\gamma$ decay. In
this case, one should consider the expressions for the third component
of the axial current,
\begin{equation}
  \partial^{\mu}(\bar{u}\gamma_{\mu}\gamma_5
  u-\bar{d}\gamma_{\mu}\gamma_5 d) =\frac{e^2}{8\pi^2}
  F\tilde{F},
\end{equation}
and identify the corresponding transformation parameter $\alpha_{3}$
with Goldstone mode $\pi^0$ such that the effective Lagrangian takes
the familiar form
\begin{equation}
  \delta \Lagrangian_{\pi^0\gamma\gamma}
  = \frac{e^2}{8\pi^2}
  F\tilde{F}
  \frac{\pi^0}{f_{\pi}\sqrt{2}}.
\end{equation}

Now, we want to derive a similar expression for the effective
Lagrangian describing interaction of the light $\eta'$ field with
electromagnetism in CFL and 2SC phases\footnote{Anomalous
  electromagnetic interaction for the $SU(3)$ Goldstone modes in the
  CFL phase was discussed previously using a different
  approach~\cite{Nowak:2000wa}.  As far as we know, the anomalous
  electromagnetic interaction with the singlet $\eta'$ phase has not
  been previously discussed in the literature.}  As we mentioned
above, the operator expression for the anomaly~(\ref{eq:anomaly})
remains unchanged: the singlet phase $\alpha$ defined
in~(\ref{eq:Atransformation}) is identified with the physical
Goldstone mode $\eta'$, which is now the phase of the diquark
condensate instead of the chiral condensate
\begin{align}
  \label{5}
  \Sigma &\sim e^{-i\varphi_A}|\Sigma|, & 
  \alpha &= \frac{\varphi_A}{2}
  =\frac{\eta'}{2f_A}.
\end{align}
This is not the end of the story however, because, as it has been
known since~\cite{Alford:1999pb}, in dense QCD matter, the electromagnetic
field strength $F_{\mu\nu}$ and the electric charge $e$ are not the
appropriate physical quantities.  Rather, a combination of the
electromagnetic field $A_{\mu}$ with $8^{\text{th}}$ component of the
gluon field $A_{\mu}^8$ acts as a physical massless photon
$\mathcal{A}$ field~\cite{Alford:1999pb}:
\begin{subequations}
  \label{eq:PhysicalPhoton}  
  \begin{align}
    \mathcal{A}_{\mu} &=  A_{\mu}\cos\theta+  A_{\mu}^8\sin\theta \\
    \mathcal{A}^8_{\mu} &= - A_{\mu}\sin\theta +  A_{\mu}^8\cos\theta \\
    \cos\theta &= \frac{g}{\sqrt{g^2+4\eta^2 e^2}} \\
    \sin\theta &= \frac{2e\eta}{\sqrt{g^2+4\eta^2 e^2}} \\
    \mathfrak{e} &= \frac{e g}{\sqrt{g^2+4\eta^2 e^2}},
  \end{align}
\end{subequations}
where the $\mathcal{A}$ is the physical ``photon'' field and
$\mathfrak{e}$ is the physical charge.  The values for
parameter $\eta$ entering the expressions (\ref{eq:PhysicalPhoton})
are given by~\cite{Alford:1999pb},
\begin{align}
  \label{7}
  \eta_{CFL} &= \frac{1}{\sqrt{3}}, &
  \eta_{2SC} &= -\frac{1}{2\sqrt{3}}.
\end{align}
We note that our expression for the angle $\theta$ in terms of $e$ and $g$
is obtained from the one presented in~\cite{Alford:1999pb} by changing
$g \rightarrow g/2$.  The difference is due to the non-standard
definition of the strong coupling constant $g$ in~\cite{Alford:1999pb}
(the absence of the factor $1/2$ in front of $g$ in the covariant
derivative in~\cite{Alford:1999pb}).  We use the standard definition
for the strong coupling constant such that the chiral anomaly is given
by~(\ref{eq:anomaly}).  Using our normalization for the $\eta'$ field
and expressing the anomaly in terms of the physical electromagnetic
field $\mathcal{F}_{\mu\nu}$ we arrive to the following effective
Lagrangian describing the interaction of the $\eta'$ with the
electromagnetic fields in the CFL phase:
\begin{equation}
  \label{eq:deltaLCFL}
  \delta \Lagrangian^{(\text{CFL})}_{\eta'\mathcal{F}\tilde{\mathcal{F}}}
  =\frac{\mathfrak{e}^2}{4\pi^2} 
  \mathcal{F}\tilde{\mathcal{F}}\frac{\varphi_A}{2}.
\end{equation}
It is interesting to note that half of this result is due to the
original electromagnetic interaction, $\sim F\tilde{F}$ while the
other half is due to the gluonic part of the
anomaly~(\ref{eq:anomaly}).  One can repeat the same calculations for
the 2SC phase with the result similar to~(\ref{eq:deltaLCFL}):
\begin{equation}
  \label{eq:deltaL2SC}
  \delta \Lagrangian^{(\text{2SC})}_{\eta'\mathcal{F}\tilde{\mathcal{F}}}
  =\frac{\mathfrak{e}^2}{8\pi^2} 
  \mathcal{F}\tilde{\mathcal{F}}\frac{\varphi_A}{2}.
\end{equation}
However, for the 2SC phase, only one sixth of this contribution is due
to the original gluon term in the anomaly.  One should also note that
the same procedure determines the anomalous coupling constants $\sim
g_{\eta'\mathcal{F}\mathcal{G}}$ describing the decay of a heavy gluon
to an $\eta'$ and a photon---the decay similar to
$\rho\rightarrow\pi\gamma$ in usual QCD.  We shall not discuss further
this physics involving the heavy particle ($m_G\sim \Delta$) in the
present paper.

\subsection{Superconducting strings}
\label{sec:superc-strings}
In this section we wish to analyze the system of light particles
$\eta'$ (the phase $\varphi_A$) and photon $\mathcal{F}_{\mu\nu}$.
Combining (\ref{eq:LeffA}) and (\ref{eq:deltaLCFL}) we have the
following effective Lagrangian:
\begin{widetext}
  \begin{equation}
    \label{eq:LeffEM}
    \Lagrangian_{\text{eff}} = f^2 \left((\partial_0\varphi_A)^2 
    - u^2 (\partial_i\varphi_A)^2\right) 
    -V_{\mathrm{inst}}(\varphi_A)
    -\frac{1}{4}\mathcal{F}_{\mu\nu}\mathcal{F}^{\mu\nu}
    +\frac{\mathfrak{e}^2}{8\pi^2}\left(\frac{\varphi_A}{2}\right)
    \epsilon_{\mu\nu\lambda\sigma}\mathcal{F}^{\mu\nu}\mathcal{F}^{\lambda\sigma}.
  \end{equation}
\end{widetext}
We approach this problem for large $\mu$. In this case, as a first
approximation, we can omit the instanton contribution
$V_{\mathrm{inst}}$ as we did for the discussion of a single string.
This allows us to treat the problem of string-electromagnetism exactly
at large distances.  If we turn on the small instanton term and
explicitly break the $U(1)_A$ symmetry, then domain walls are also
allowed and the problem becomes much more complicated. We speculate on
a possible effects of these physically relevant domain walls in the
conclusion.

The effective theory~(\ref{eq:LeffEM}) is almost identical to that
studied in~\cite{Kaplan:1988kh}.  We begin our analysis by considering
a single spinning global axial string lying along the $z$-axis and use
cylindrical coordinates $(r,\phi)$:
\begin{equation}
  \label{eq:phi}
  \Phi(t,r,\phi) = \rho(r)e^{i\varphi_A(t,\phi)}.
\end{equation}
The radial solution $\rho(r)$ was presented in
Figure~\ref{fig:string}, but we will only be concerned with distances
large with respect to the core size of the string $r \gg r_c \sim
\Delta$.  In this regime, the effective theory~(\ref{eq:LeffEM})
becomes valid and we have the following equations of motion for the
``electromagnetic'' field:
\begin{equation}
  \partial_\mu\mathcal{F}^{\mu\nu} 
  = \frac{\mathfrak{e}^2}{2\pi^2}
  (\partial_\mu\varphi_A)\tilde{\mathcal{F}}^{\mu\nu}
  = \frac{2\alpha}{\pi}
  (\partial_\mu\varphi_A)\tilde{\mathcal{F}}^{\mu\nu}
\end{equation}
where $\alpha = \mathfrak{e}^2/4\pi$ is the modified fine structure
constant and the background field $\varphi_A$ has the solution
\begin{equation}
  \varphi_A(t,\phi) = \dot{\varphi}_A t + \phi
  \quad\text{ where }\quad
  \phi = \tan^{-1}\frac{y}{x}.
\end{equation}
We assume that the rate at which the string is spinning is small:
\begin{equation}
  \dot{\varphi}_A \equiv \frac{\partial\varphi_A}{\partial t} \ll \Delta^{-1}.
\end{equation}
The origin for this assumption is that, while $\varphi_B$ strings
have very high frequencies of order of $\mu$, (\ref{eq:Bw}), the
$\varphi_A$ strings do not spin on their own.  They can, however,
receive some angular momentum through the interaction with the
surrounding medium an/or the $\varphi_B$ strings.  The interaction of
the Goldstone fields with other particles is expected to be suppressed
by some power of $\mu^{-1}$.  Therefore, we expect $\dot{\varphi}_A$
to be small with respect to all relevant scales.

The anomalous Maxwell's equations~(\ref{eq:LeffEM}) in the presence of
a time dependent global string background have the following form far
from the core of the string ($r\gg \Delta^{-1}$)
\begin{subequations}
  \begin{align}
    \label{11}
    -\frac{\partial \mathcal{E}_r}{\partial t}
    +\left(\frac{1}{r}\frac{\partial \mathcal{B}_z}{\partial \phi}
      -\frac{\partial \mathcal{B}_{\phi}}{\partial z}\right)
    &= -\frac{2\alpha}{\pi r}\mathcal{E}_z
    -\frac{2\alpha\dot{\varphi}_A}{\pi}\mathcal{B}_r,\\
    -\frac{\partial \mathcal{E}_z}{\partial t}
    +\left(\frac{1}{r}\frac{\partial (r\mathcal{B}_{\phi})}{\partial r}
      -\frac{1}{r}\frac{\partial \mathcal{B}_{z}}{\partial \phi}\right)
    &=\frac{2\alpha}{\pi r}\mathcal{E}_r-\frac{2\alpha\dot{\varphi}_A}{\pi}\mathcal{B}_z,\\
    -\frac{\partial \mathcal{E}_{\phi}}{\partial t}
    -\frac{\partial \mathcal{B}_z}{\partial r}
    +\frac{\partial \mathcal{B}_{r}}{\partial z}
    &=
    -\frac{2\alpha\dot{\varphi}_A}{\pi}\mathcal{B}_{\phi}, \\
    \frac{1}{r}\frac{\partial(r\mathcal{E}_{r})}{\partial r}
      +\frac{\partial \mathcal{E}_{z}}{\partial z}
      +\frac{1}{r}\frac{\partial \mathcal{E}_{\phi}}{\partial \phi}
      &=\frac{2\alpha}{\pi r}\mathcal{B}_{\phi}.
  \end{align}
\end{subequations}
In the limit of $\dot{\varphi}_A\rightarrow 0$, these results reduce
to those presented in~\cite{Kaplan:1988kh} with the replacement
$2\alpha \rightarrow \alpha$ which is due to the extra factor $2$ in
our expression~(\ref{eq:deltaLCFL}) for the anomaly as compared with
the axion model considered in~\cite{Kaplan:1988kh}.

We were not able to find a complete solution of these linear equations
over all of space, however, a $z$-independent solution for radii
larger than the core size but not too large $\Delta^{-1} \sim r_c \ll
r \ll \dot{\varphi}_A^{-1}$ can be easily found:
\begin{subequations}
  \label{eq:solEM}
  \begin{align}
    \mathcal{E}_{r} &= C_+r^{-1+2\alpha/\pi}+C_-r^{-1-2\alpha/\pi},  
    \label{eq:solEMa}\\
    \mathcal{B}_{\phi} &= C_+r^{-1+2\alpha/\pi}-C_-r^{-1-2\alpha/\pi},
    \label{eq:solEMb}\\
    \mathcal{B}_{z} &= \dot{\varphi}_A 
    (C_+r^{2\alpha/\pi}+C_-r^{-2\alpha/\pi}),
    \label{eq:solEMc}
  \end{align}
\end{subequations}
with the other field components vanishing.  Coefficients $C_{\pm}$
need to be determined by matching the solution~(\ref{eq:solEM}) with
the behaviour of fields in the core region $r\simeq \Delta$ where our
effective Lagrangian approach breaks down.  In what follows, we
estimate these coefficients using dimensional arguments.

The solution~(\ref{eq:solEM}) is reduced to the corresponding
expression~\cite{Kaplan:1988kh} in the limit $\dot{\varphi}_A=0$.  In
this case, the solution becomes valid for arbitrarily large $r$.  Our
approximate solution~(\ref{eq:solEM}) is limited to the range
$\Delta^{-1}\ll r\ll \dot{\varphi}_A^{-1}$.  At shorter ranges we must
further develop our microscopic model.

The most important feature of the solution~(\ref{eq:solEM}) is that,
for a given $C_+\neq 0$ or $C_-\neq 0$, the solution has the Lorentz
property $\mathcal{B}_{\phi} =\pm \mathcal{E}_r$---as if the string
carries a light-like current four vector with current density along
the $z$-axis $\mathfrak{j}_{\mu}= (\mathfrak{j}, 0,0,\pm
\mathfrak{j})$ with the whole system spinning with small frequency
$\dot{\varphi}_A\ll 1$.  The presence of a small $\mathcal{B}_z$ field
in (\ref{eq:solEMc}) which is due to the spinning of the static
solution can be easily understood by boosting to the local frame
moving with the speed $v_{\phi} =- \dot{\varphi}_A r$ at the point
$r$. 

We should note here, that the presence of a current with the property
$\mathfrak{j}_{\mu} = (\mathfrak{j}, 0,0,\pm \mathfrak{j})$ in the
system defined by~(\ref{eq:LeffEM}) was established by analyzing the
only large distance physics without formulating a microscopic model of
the core.  A similar result was obtained in~\cite{Kaplan:1988kh}.
There, however, the effective theory was taken to be exact, allowing
for a detailed analysis of the core physics.  In that case, it can be
seen that such a current results from a charged Dirac fermionic zero
mode present in the string background
(see~\cite{Weinberg:1981eu,Witten:1985eb,Callan:1985sa,Jackiw:1981ee}
for more details).  Unfortunately, in our case, the effective
theory~(\ref{eq:LeffEM}) breaks down at short distance scales: we only
have theoretical control over the large distance physics, however we
believe that the microscopical explanation of the property
$\mathfrak{j}_{\mu} = (\mathfrak{j}, 0,0,\pm \mathfrak{j})$ in our
case is very similar to the microscopical explanation given
in~\cite{Kaplan:1988kh}: namely, that it is due to zero modes of the
charged fermion field which travels inside the core of the string.

The presence of zero modes (localized in the $(xy)$ direction) in the
string background is a very general property of such a background and
is a trivial consequence of an index
theorem,~\cite{Weinberg:1981eu,Witten:1985eb,Callan:1985sa,Jackiw:1981ee}.
Although this topic is beyond the scope of the present paper, we would
like to note that, in the colour superconducting CFL or 2SC phases,
the interaction between the gapped fermions close to the Fermi surface
and the diquark condensate has a more complicated algebraic structure
than the simple Yukawa coupling of a single fermion considered
previously~\cite{Witten:1985eb,Callan:1985sa,Kaplan:1988kh}.
Nevertheless, one can explicitly demonstrate the presence of the
fermionic zero modes which would have been the regular gapped
excitations in the absence of the string background.  Therefore, we
believe that the microscopical explanation of the result
$\mathfrak{j}_{\mu}=(\mathfrak{j}, 0,0,\pm \mathfrak{j})$ in our case
is analogous to the explanation given in~\cite{Kaplan:1988kh}.
However, unlike the case in~\cite{Kaplan:1988kh}, we expect that both
the coefficients $C_{\pm}$ are nonzero due to the lack of Lorentz
invariance in our system. (The dispersion laws for in-medium fermions
do not have a standard Lorentz-invariant form.)

One of the consequences of localized charged fermionic zero modes in a
string background is superconductivity. Indeed, as was demonstrated
in~\cite{Witten:1985eb}, the problem can be reduced to a two
dimensional effective theory which describes massless charged fermions
(our original zero modes in four dimensions), and photons (our
physical ``photon'' fields $\mathcal{A}_{\mu}$) which are coupled to
the massless fermions through the physical charge $\mathfrak{e}$.  It
was also demonstrated in~\cite{Witten:1985eb} that such a system
describes a superconducting string in the sense that, if external
electric field is applied along the $z$ direction, it results in
persistent current along the string.  We shall not repeat all these
well known arguments, which are based on the bosonization technique of
the (effectively two-dimensional) localized zero modes.  Instead, we
refer the reader to the original paper~\cite{Witten:1985eb}.

Thus, we can only present a dimensional estimate at this time: For a
given orientation of the string solution $\varphi_A(r,\phi)= +\phi$ we
expect that both $C_\pm$ of Equation~(\ref{eq:solEM}) are of the same
order and can be estimated dimensionally to be
\begin{align}
  C_- &\sim \mathfrak{e}\Delta (\Delta)^{-2\alpha/\pi}, &
  C_+ &\sim \mathfrak{e}\Delta (\Delta)^{+2\alpha/\pi}.
\end{align}
This result behaves in the same way as if fermions trapped in
transverse zero modes travel in opposite directions (without
cancellation, $C_+\neq C_-$) at the speed of light.  We should
remark that, due to the linearity of Maxwell's equations, a linear
superposition of two sources $c_+(\mathfrak{j}, 0,0,
+\mathfrak{j})+c_-(\mathfrak{j}, 0,0, -\mathfrak{j})$ produces a
desirable superposition of the corresponding solutions
(\ref{eq:solEM}). Such  electromagnetic behaviour of the string 
is reminiscent of the behaviour of free superconducting
strings~\cite{Witten:1985eb,Aryal:1987pw}. 

The question of how the external electric  field
may be generated is a different question
which is not addressed in the present paper. However, 
we should remark here  that, in the core of a neutron
star, the  current could be  generated by
 the motion of the strings through external magnetic fields already  present
in neutron star~\cite{Witten:1985eb}.
 
To conclude this section, we estimate the maximum current which can be
achieved in a colour superconductor.  Under an external electric field
applied parallel to the axis of the string, the fermion current will
grow.  Eventually, however, it will saturate: If the mass of the
fermions away from the string is $\Delta$, then once the Fermi
momentum of the charge carriers rises above $p_F > \Delta$, it will be
energetically favourable for the fermion to leave the string.
Therefore, one expects that the maximum current supported by a single
string would be of the order
\begin{equation}
  \mathfrak{j}_{\text{max}} \sim
  \frac{\mathfrak{e}\Delta}{2\pi}
\end{equation}
which cannot exceed $2\cdot 10^3$ A.

 We do not expect, however, that
a single, infinitely long string is realized in nature. Rather, we
expect that the strings will take the shape of a ring or organize into
more complicated objects.
We presently cannot say more about the results of the rather
complicated dynamics of strings in dense quark matter in the presence
of strong external magnetic fields. However, we refer to some of the
results  discussed in~\cite{Witten:1985eb} and 
 the textbook~\cite{Vilenkin:1994}.
As a first step to understanding the dynamics of this complicated system,
one has to understand the forces which act between the strings.  We
discuss this subject in the Conclusion.

\section{Conclusion}
\label{sec:conclusion}
In this paper, we have considered the existence of classically stable global
string configurations in high density QCD.  We have shown that, in the
$N_f=3$ colour flavour locking (CFL) phase, baryonic strings resulting
from the $U(1)_B$ symmetry of the condensate can carry angular
momentum.  A similar situation arises for $U(1)_S$ strings 
if  a kaon condensate forms.  In addition, we have shown
that the axial strings resulting from the approximate $U(1)_A$
symmetry may exist in both the $N_f=3$ CFL phase and the $N_f=2$ (2SC)
phase.

Even though the fields that form these global strings are neutral, we
show that the $U(1)_A$ strings have non-trivial electromagnetic
properties due to the anomalous coupling between the $\eta'$ field and
a massless ``photon'' that is a mixture of the eighth gluon and photon
fields.  Due to this anomalous interaction, axial strings behave as
superconducting string.

As we have already mentioned, however, in general,
strings will not be formed as infinitely long
strings in isolation. The details of the interaction between strings
is quite varied and complicated: stable structures like helices, rings
and vortons may form, or the strings may form networks or tangle.
Besides that, $U(1)_A$ strings as well as $U(1)_S$ strings will be
bounded by domain walls which themselves can decay into
string-antistring configurations.  In order to understand the possible
dynamics of the system one must know different forces which determine
the dynamics.

In this Conclusion we shall only comment on a few simple possibilities
here and refer the reader to the textbook~\cite{Vilenkin:1994} for a
discussion about general string dynamics.  We shall consider the
following interactions between strings:
\begin{itemize}
\item Inter-Vortex Force: $F_{VV}$.
\item Domain Wall Force: $F_{DW}$.
\item Anomalous Electromagnetic Force: $F_{EM}$.
\end{itemize}
We shall briefly summarize these forces and provide qualitative
estimates in terms of the parameters of the QCD phase.  In the
following we assume that two infinitely long global strings lie
parallel to the $z$-axis with separation $d$.  We label the strings by
their topological ``charge'' $n$ which corresponds to their winding
number~(\ref{eq:n}) with respect to the positive $z$ axis.  Strings of
the same charge have the same orientation: String of opposite charge
can annihilate.  All forces are presented as a force per unit length
of the string.

\paragraph{Inter-Vortex Force}
The most important and well-known interaction is due to the exchange
of massless Goldstone particles.  The force is estimated by
considering the energy of a configuration of two strings.  This
approximation is valid for distances $d$ much larger than the core
size $d \gg
\Delta^{-1}$~\cite{Shellard:1987bv,Perivolaropoulos:1992du}
\begin{equation}
  \label{a1}
  F_{VV} \sim \mp\frac{4\pi f^2}{d}.
\end{equation}
This force is repulsive for strings with the same charge, and
attractive for strings of opposite charge. Physically, strings of
opposite charge tend to attract (energetically they try to annihilate
and restore the vacuum everywhere) while strings of the same charge
repel: strings of higher charge $n>1$ will split into several strings
of single winding which will tend to move away from each other.

For baryonic $U(1)_B$ strings this result extends to arbitrarily large
distances $d$ because the Goldstone bosons are massless.  For $U(1)_S$
strings, this description is only valid for distances $d <
m^{-1}\simeq(50\text{ KeV})^{-1}$ because the Goldstone bosons are in
fact massive due to weak interaction~\cite{Son:2001xd}.  A similar
situation occurs for the axial $U(1)_A$ strings where this
description is only valid for distances $d < m_{\eta'}^{-1}$ because
the Goldstone bosons is massive due to instanton
effects~(\ref{eq:meta}).  Beyond this range, this inter-vortex force
falls away and the domain wall force starts to dominate.

\paragraph{Domain Wall Force}
On distance scales $d > m^{-1}$, where $m=m_{\eta'}$ for $U(1)_A$
strings and $m=m_{\varphi_S}\sim \sqrt{a_S}\Delta$ for $U(1)_S$
strings, (\ref{eq:K_0}), one will see that the $U(1)_A$ and $U(1)_S$
strings are really embedded in domain walls of thickness $m^{-1}$.
The scale for this interaction is set by the corresponding domain wall
tension $\sigma$ calculated for $\varphi_A$ field in~\cite{Son:2000fh}
and for $\varphi_S$ field in~\cite{Son:2001xd}:
\begin{equation}
  \label{eq:FDW}
  F_{DW} \sim \sigma \sim f^2 m .
\end{equation}
This force can be attractive for opposite charge strings if a domain
wall connects them.  Domain walls are not formed between strings of
the same charge if we neglect the process of nucleation which is
equivalent to the creation of the string-antistring pairs.

\paragraph{Anomalous Electromagnetic Force}
For distances between $\Delta^{-1}\ll d \ll \dot{\varphi}_A^{-1}$ the
axial string behave as current carrying wires.  Through the exchange of
``photons'', these strings thus interact with force
\begin{equation}
  \label{eq:FEM}
  F_{EM} \sim \frac{\mathfrak{j}\mathfrak{j}}{d} 
  \sim \pm\frac{\alpha}{\pi}\frac{\Delta^2}{d}.
\end{equation}
This force is attractive for strings of the same charge and repulsive
for strings of opposite charge.  Physically, strings carrying current
in the same direction want to bunch together into a ``wire''.  In
comparison with the other two forces, however, the anomalous
electromagnetic force is suppressed by a factors of $\alpha$ and
$\Delta/\mu$ and can be neglected in comparison with these other
forces. 

Besides that, when a $U(1)_A$ string moves at velocity $v$ across an
external magnetic field, the effective superconducting current should
build at a very high rate, (see~\cite{Witten:1985eb}).  In such a
situation, an electric field is also generated leading to the creation
of particles, a process which we do not discuss here.
 
One more effect which we would like to mention here and which deserves
for a further study is the binding strings by the domain wall force
(\ref{eq:FDW}).  In a situation similar to the $N=1$ axion
model~\cite{Vilenkin:1982ks} such a configuration decays very quickly.
In our case when $U(1)_A$ string is superconducting and can form a
ring with a dragged magnetic flux crossing the ring, the situation may
not be so obvious. One should better understand the system of
superconducting string with attached domain wall in the dense
background in the presence of a nonzero magnetic field, before one can
make any conclusions regarding the system.  We suspect that due to the
interactions which were not present in the axion case, some stable
configurations (like vortons) will be possible.
 
The most likely place to encounter bulk high-density superconducting
quark matter is in the core of neutron stars.  Typically, neutron
stars are rapidly spinning, and we suspect that during the formation
of the core, this rotation would impart some angular momentum to the
superconducting core.  As we showed in Section
\ref{sec:spinning-strings}, spinning global $U(1)_B$ and $U(1)_S$
strings carry angular momentum that is correlated with the string's
charge.

It remains to be seen whether these strings can have any impact
on observable effects (they might, for example, affect glitches:
sudden increases in the rotation frequency $\omega$ of neutron stars
by as much as $\Delta\omega/\omega\sim 10^{-6}$, or the magnetic field
structure and evolution).

There will likely be many other effects due to the interactions and
electromagnetic properties of global $U(1)$ strings that will play a
role in the physics of superconducting phases of high density quark
matter.  These effects might be closely related to analogous effects
studied for cosmic strings (we
 refer the reader to the
textbook~\cite{Vilenkin:1994}), and studied in condensed matter
physics.  We hope that the present paper will initiate some activity
in this direction.

\begin{acknowledgments}
  We are indebted to A.~Kovner, D.T.Son, M.~Stephanov, T.~Sch\"afer,
  E.~V.~Shuryak and, K.~Zarembo for discussions regarding different
  aspects of the global strings at high density in QCD.  MMF would
  also like to thank F.~Wilczek for useful discussions.  AZ is
  supported, in part, by the NSERC of Canada.  MMF is supported by a
  Presidential Fellowship from MIT.
\end{acknowledgments}

\ifrevtex{}{\bibliographystyle{s-JHEP}}

\bibliography{master}
\end{document}

%%% Local Variables: 
%%% mode: latex
%%% TeX-master: t
%%% End: 